\documentclass{webofc}
\usepackage[varg]{txfonts}   
\usepackage{hyperref}
\usepackage{url}
\hypersetup{colorlinks=true,citecolor=blue,urlcolor=blue,linkcolor=blue}
\begin{document}

\def\bear{\begin{eqnarray}}
\def\ear{\end{eqnarray}}
\def\e{{\rm e}}

\title{Pair creation in electric fields}
%
%

\author{\firstname{Christian} \lastname{Schubert}
\inst{1}
\fnsep\thanks{\email{christianschubert137@gmail.com}} }

\institute{Facultad de Ciencias Físico-Matemáticas, Universidad Michoacana de San Nicolás de Hidalgo, Avenida Francisco J. Mújica, 58060 Morelia, Michoacán, Mexico}

\abstract{%
Sauter-Schwinger pair creation in electromagnetic fields is a fundamental prediction of QED and one of the motivations for the present efforts in constructing super-strong lasers. I will give a historical review of the subject, and then focus on two recent developments. The first one is the worldline instanton formalism, a sophisticated version of the WKB approximation that makes it possible to calculate the pair creation rate for complicated field configurations. The second one is an adaptation of the 
Dirac-Heisenberg-Wigner formalism suitable for a detailed study of the formation of real particles in time and space. 
}
\maketitle

\setcounter{footnote}{1} 
\section{Historical review}
\label{intro}

\begin{figure}[h]
\centering
\includegraphics[width=5cm,clip]{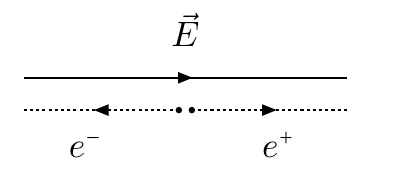}  
\caption{Vacuum tunneling of an electron-positron pair in an electric field.}
\label{fig-pairtunnel}       
\end{figure}

In his famous 1931 paper ``\"Uber das Verhalten eines Elektrons im homogenen elektrischen Feld nach der relativistischen Theorie Diracs
\footnote{On the behavior of an electron in a homogeneous electric field according to Dirac's theory.}" \cite{sauter} Sauter solved the Dirac equation
in a constant electric field and found a certain probability for ``a transition from positive to negative impulses''. This was later to be interpreted as electron-positron pair
creation by the field, or, from a modern field-theory point of view, as ``vacuum tunneling'' of a virtual to a real pair: 
due to a statistical fluctuation governed by the time-energy uncertainty relation, 
a virtual pair separates out far enough to draw its rest mass energy from the field
and become real (Fig. \ref{fig-pairtunnel}).

Twenty years later, Schwinger \cite{schwinger51} computed the pair-production rate $P$ for the constant-field case using his novel
effective action techniques. Assuming that the rate is small, it can be approximately computed from the 
imaginary part of the action,
\bear
P \approx 2 {\rm Im}\Gamma (E) \, .
\label{Papprox}
\ear
In the constant-field case, the effective action is just the space-time integral of the
Euler-Heisenberg Lagrangian ${\cal L}_{EH}(E)$, obtained by Heisenberg and Euler in 1936 \cite{eulhei},
for whose imaginary part Schwinger found the well-known expansion
\begin{eqnarray}
{\rm Im} {{\cal L}_{EH}}(E) &=&  \frac{m^4}{8\pi^3}
\beta^2\, \sum_{k=1}^\infty \frac{1}{k^2}
\,\exp\left[-\frac{\pi k}{\beta}\right]
\label{schwinger}
\end{eqnarray}
($\beta = eE/m^2$). We note that

\begin{itemize}

\item
${\rm Im}{\cal L}(E)$ depends on $E$ non-perturbatively, which lends support to the tunneling picture. 

\item
The total pair creation rate is given by the leading $k=1$ term (seeing this requires an analysis of the corrections to \eqref{Papprox}). 

\item
The $k\geq 2$ terms carry information on the coherent creation of $k$ pairs in one Compton volume. 

\item
The corresponding formula for scalar QED differs from \eqref{schwinger} only by the normalization and sign changes
from the difference between Bose-Einstein and Fermi-Dirac statistics,
\begin{eqnarray}
{\rm Im} {{\cal L}_{\rm scal}}(E) &=&  \frac{m^4}{16\pi^3}
\beta^2\, \sum_{k=1}^\infty \frac{(-1)^{k+1}}{k^2}
\,\exp\left[-\frac{\pi k}{\beta}\right]
\, .
\label{schwingerscal}
\end{eqnarray}

\end{itemize}

For a constant field the pair creation rate is exponentially small for field strengths below 
the critical field strength $E_{\rm crit}$, 

\bear
E_{\rm crit} = \frac{m^2}{e} \approx 10^{18} {\rm V}/{\rm m} \, .
\ear
This critical field strength is such that an electron will collect its
rest energy from the field on a distance of one Compton wave length. 
Present-day lasers are still two to three orders of magnitude away from this (for reviews of the experimental situation, see \cite{LUXE,FACET,fikkstt}). 
To have any chance at seeing pair creation soon, complicated laser configurations
will have to be used to lower the pair creation threshold. To mention here just two of the many
proposals that have been put forward over the last fifteen years, counterpropagating 
linearly polarized lasers were proposed by Ruf et al. \cite{rmmhc} and superimposing a plane-wave X-ray beam with a strongly 
focused optical laser pulse by Dunne et al. \cite{dugisc} (for an overview see \cite{fikkstt}). 

\section{Approximation methods for Schwinger pair creation}

For the field configurations corresponding to most of these proposals an exact calculation of the
pair-creation rate is out of the question; reliable approximation methods are called for.
Until the eighties, virtually the only method available in this context was WKB \cite{keldysh65,breitz70,narnik70,popov72}.
A more sophisticated version of WKB, better adapted to the relativistic nature of the pair-creation process, 
is the {\it worldline instanton formalism}. It was introduced by Affleck et al. \cite{afalma} in 1982 for scalar pair-production in
a constant field, but gained popularity only following its generalization to fermions and non-constant fields by G.V. Dunne
and the author \cite{63}. 

For purely time-dependent fields, the {\it quantum kinetic approach} was developed in the early nineties,
based on a { Vlasov-type equation} \cite{kescm1,kescm2,sbrspt,healgi}. 

Even more recently, it has been found that the Dirac-Heisenberg-Wigner formalism, invented by Wigner in 1932 \cite{wigner32} 
and further developed in \cite{vagyel,bigora,ochhei} can provide a more detailed picture of pair creation process
\cite{healgi2,hebenstreit,ilmaza,kohlfuerst,dialko}. 

In the following, we will discuss each of these three approaches in turn. 

\section{Worldline instantons}

The worldline instanton formalism is based on Feynman's ``worldline representation'' of the QED effective action.
For scalar QED it reads \cite{feynman1950}

\bear
\Gamma_{\rm scal} [A] &=&
\int_0^{\infty}{dT\over T}\, \e^{-{m^2}T}
\int {\cal D}x(\tau) 
\, \e^{-S[x(\tau)]}
\, ,
\nonumber\\
S[x(\tau)] &=& 
\int_0^Td\tau 
\Bigl({\dot x^2\over 4} +ieA\cdot \dot x \Bigr)
\, .
\nonumber\\
\ear
Here $ m$ and $ T $ are the mass and proper time of the loop scalar,
and the path integral $ \int{\cal D}x(\tau)$
is over closed trajectories in  Euclidean  spacetime. 

In 1982, Affleck, Alvarez and Manton \cite{afalma} used this representation for an elegant
rederivation of Schwinger's formula \eqref{schwingerscal} by simply replacing the 
path integral by a single stationary trajectory, the worldline instanton. For a constant electric field pointing into
the $z$ direction, ${\vec E} = (0,0,E) = const.$, this trajectory is given simply by a circle in the $z-t$ place,
\bear
x^{\rm cl}_k(u) &=& {m\over eE}\,\Bigl(x_1,x_2,{\rm cos}(2k\pi u),{\rm sin}(2k\pi u)\Bigr)
\label{wlinst}
\ear
and it carries an index $k$ for the number of times that the circle is traversed. 
The worldline action evaluated on the instanton reproduces the exponent of the $k$th
term in \eqref{schwingerscal}, 
\bear
S[x^{\rm cl}] &=& k\pi\, {m^2\over eE}
\ear
and the prefactor determinant can (with a bit more of work) shown to provide the correct normalization. 

While in the constant-field case this method provides the exact answer (because here the path integral is gaussian),
for arbitrary electric fields it has the character of a semi-classical approximation, closely related to the 
WKB approximation \cite{kimpage1,kimpage2,kimpage3}. It can be summarized in the remarkably simple formula
\bear
{\displaystyle \int_{x(T)=x(0)=x^{(0)}}}\hspace{-20pt}{\cal D}x(\tau)
\, \e^{-S[x(\tau)]}
\approx
\e^{i\theta}
\e^{-S[x^{\rm cl}](T)} 
\frac{1}{(4\pi T)^2}
\sqrt{
\biggl\vert {\rm det}\Bigl[\eta^{(\lambda)}_{\mu,{\rm free}}(T)\Bigr]\biggr\vert\over
\biggl\vert {\rm det}\Bigl[\eta^{(\lambda)}_{\mu}(T)\Bigr]\biggr\vert}
\ear
where the extremal action trajectory $x^{\rm cl}(u)$
is a  periodic solution of the (euclidean) Lorentz 
force equation, the $\eta_{\mu}^{(\lambda)}$ are zero modes of the Hessian fluctuation operator around it,
and the phase factor $\e^{i\theta}$ is related to the Morse index of this operator.

\begin{figure}[h]
\hspace{-35pt}
\centerline{\includegraphics[scale=.65]{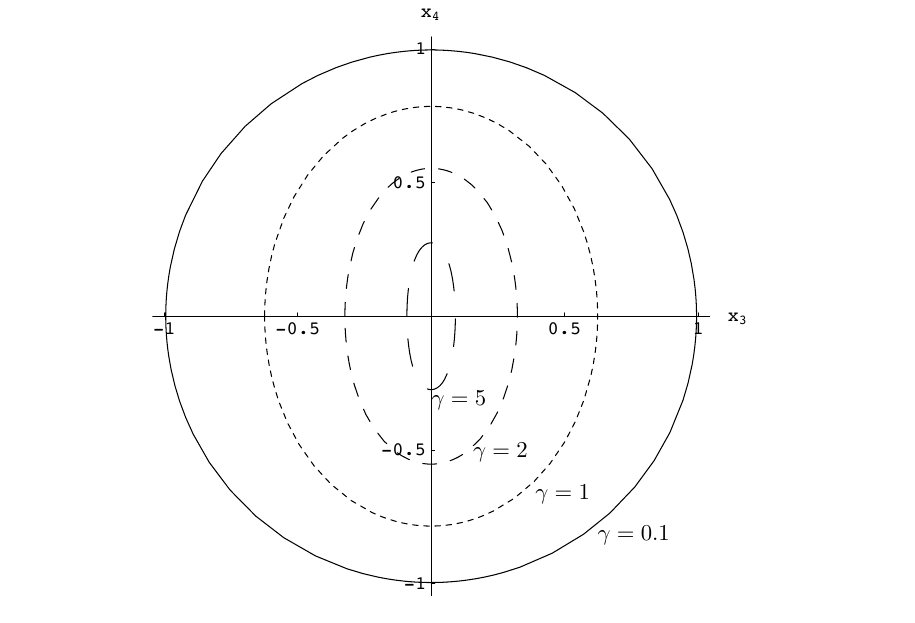}}
  \caption{Plot of the worldline instanton paths 
  in the $ (x_3,x_4)$ plane
  for $ E(t) = E{\rm sech}^2(\omega t)$. 
  The paths are shown for
  various values of the adiabaticity parameter $ \gamma$. 
  $ x_{3,4}$ have been expressed
  in units of $\frac{m}{eE}$.}
 \label{fig-wlinst_sauter(t)}
 \vspace{-20pt}
\end{figure}

For the well-studied cases of a time-like or space-like {\it Sauter field} the worldline instantons can still be
given in closed form. The time-dependent Sauter case, defined by the two-parameter single-bump field
$E(t)=E\, {\rm sech}^2(\omega\,t)$, has the simple trigonometric instanton solution 
\bear
x_{k3}^{\rm cl}(u)&=&-\frac{1}{\omega}\,\frac{1}{\sqrt{1+\gamma^2}} \, {\rm arcsinh}\left[\gamma\, \cos\left(2 k\pi u\right)\right]\nonumber\\
x_{k4}^{\rm cl}(u)&=&\frac{1}{\omega}\, \arcsin\left[\frac{\gamma}{\sqrt{1+\gamma^2}}\, \sin\left(2 k \pi \, u\right)\right]
\nonumber\\
\ear
conveniently written in terms of the ``adiabaticity parameter'' $\gamma\equiv \frac{m\omega}{eE}$. In Fig. \ref{fig-wlinst_sauter(t)} we plot them
for various values of this parameter. 

We note that the instanton exists for any positive $\gamma$. In the limit $\gamma \to 0$ it turns into the circular one of the constant-field case above,
while with increasing $\gamma$ it shrinks and becomes elongated in the $x_4$ direction.  
Calculating the stationary worldline action one finds
\bear
S[x_k^{\rm cl}]&=&k\, \frac{m^2 \pi}{e E}\left(\frac{2}{1+\sqrt{1+\gamma^2}}\right)
\, .
\label{wlactsautert}
\ear
The action decreases with increasing  $\gamma$, therefore the pair creation rate increases.

\begin{figure}[h]
\hspace{-9pt}
\centerline{\includegraphics[height=.55\textheight]{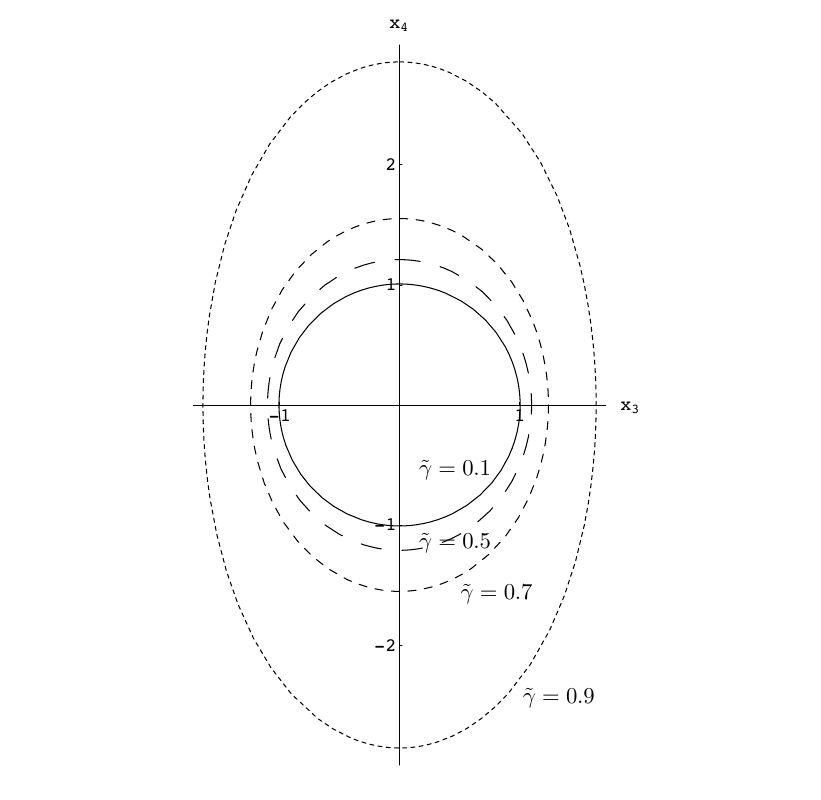}}
  \caption{Plot of the worldline instanton paths 
  in the $ (x_3,x_4)$ plane
  for $ E(x_3) = E{\rm sech}^2(x_3/d)$. 
  The paths are shown for
  various values of the parameter $\tilde \gamma$. 
  $ x_{3,4}$ have been expressed
  in units of ${m\over eE}$.}
 \label{fig-wlinst_sauter(x)}
\end{figure}

The space-like Sauter case, which we define by $E(x_3)= E {\rm sech}^2(x_3/d)$, is mathematically similar, but physically totally different.
The instanton solutions now involve hyperbolic functions:
\bear
x_{k3}^{\rm cl}(u)&=&
{m\over eE}{1\over\tilde\gamma}\,{\rm arcsinh}\biggl({\tilde\gamma\over
\sqrt{1-\tilde\gamma^2}}\,\sin(2k\pi u)\biggr)\nonumber\\
x_{k4}^{\rm cl}(u)&=&
{m\over eE}{1\over\tilde\gamma \sqrt{1-\tilde\gamma^2}}\,{\rm arcsin}
\bigl(\tilde\gamma \cos(2k\pi u)\bigr)
\ear
with an ``inhomogeneity parameter'' $\tilde \gamma\equiv \frac{m }{eEd}$. 
The trajectories are shown in Fig. \ref{fig-wlinst_sauter(x)}. 

Again they approach the constant-field instantons for small $\tilde\gamma$, but with increasing $\tilde\gamma$ they 
grow instead of shrinking. Most importantly, they cease to exist for $\tilde\gamma \ge 1$! 
A simple calculation shows that the limiting case $ \tilde\gamma = 1$ 
corresponds to the virtual particles having to run from $ x= 0$  all the way to $x=\pm \infty$ to extract their rest-mass energies from the field.
For larger $\tilde \gamma$ this becomes impossible, so that there is no pair production, no matter how much energy the field may contain; it cannot
be dispersed (note the analogy with the photoelectric effect). Thus this example provides a non-trivial prediction as well as confirmation
of the vacuum-tunneling picture. 
The instanton action is similar to \eqref{wlactsautert}, but with a crucial sign change:
\bear
S[x_k^{\rm cl}]&=&k\, \frac{m^2 \pi}{e E}\left(\frac{2}{1+\sqrt{1-\tilde\gamma^2}}\right)
\, .
\ear
\nonumber
Thus it increases with increasing $\tilde\gamma$, implying a decrease of the pair creation rate.

The Sauter field is considered a benchmark case for pair-production since there are other representations
for the exact pair-production rate suitable for numerical evaluation \cite{nikishov,giekli}. Fig. 4
(taken from \cite{64}) shows a comparison of results for the imaginary part of the effective action, normalized by the weak-field limit of the
``locally constant field approximation'', obtained by the worldline instanton approximation, a numerical evaluation of
Nikishov's representation, and a direct numerical evaluation of the worldline path integral.
Although the instanton approach has the character of a large-mass approximation, it turns out to work well in the full range
of $\tilde\gamma$, which also leaves little doubt that the vanishing of the pair-production rate for $\tilde\gamma\ge 1$ 
is an exact result and not an artefact of our semi-classical approximation. 

\begin{figure}[h]
\centerline{\includegraphics[scale=.6]{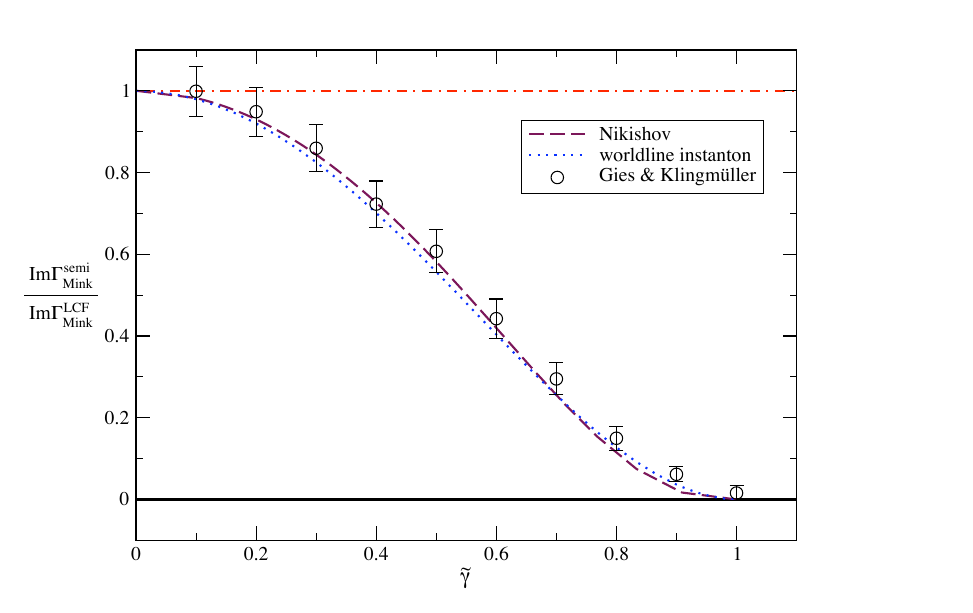}}
\label{fig-wlinstcompare}
\caption{Comparison of the imaginary part of the effective action for the space-like Sauter field
obtained by the worldline instanton method (dotted line), a numerical evaluation of
Nikishov's integral representation (dashed line), and the ``Worldline Monte Carlo" method (circles).}
\end{figure}

The worldline instanton approach has also been applied to the following combination of a strong space-like with a weak time-like Sauter field
\cite{schsch-schwinger}:
\bear
{\bf E}(t,x) = \biggl(\frac{E}{\cosh^2(kx)} + \frac{E'}{\cosh^2(\omega t)}\biggr) \, {\bf e}_x
\ear
where $ E'\ll E\ll E_{\rm crit}$. As seen in Fig. 5, for sufficiently large $\omega$ the weak temporal pulse squeezes the worldline instanton in the
$ x_0$ direction, leading to a significant enhancement of the pair creation rate.

\begin{figure}[h]
\centerline{\includegraphics[scale=1.2]{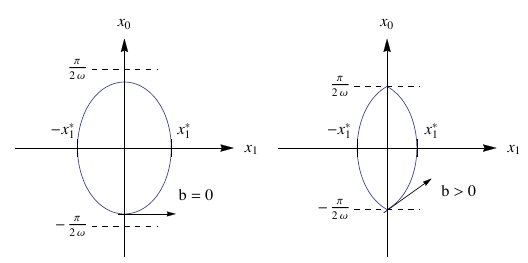}}
\caption{Worldline instanton for the combination of a space-like and a time-like Sauter field. (a) shows the instanton for $\omega < \omega_{\rm crit}$
and (b) for $\omega > \omega_{\rm crit}$ (for the definition of $\omega_{\rm crit}$ see \cite{schsch-schwinger}).} 
\label{fig-doublesauter}
\end{figure}

This example also demonstrates a feature of the worldline instanton approach that is very useful for the treatment of
complicated fields involving a superposition of several components: to see whether adding one more component will
lead to an enhancement of the pair-creation probability, in a first approximation one might just check whether it tends to
diminish or enlarge the instanton trajectory. 

Moreover, this formalism has led to the following rules \cite{63} that previously apparently were not understood in full generality: 
(i) Inhomogeneity in space tends to reduce the pair creation rate. 
An insufficiently extended field, no matter how intense, will not pair-produce.
(ii) Inhomogeneity in time tends to enhance the pair creation rate. 
A  purely time-dependent field will {\it always} give a non-zero pair creation rate.

One of the advantages of the worldline instanton formalism over the 
closely related WKB approximation is that it makes it straightforward
to incorporate loop corrections involving the internal exchange of photons. In \cite{afalma} Affleck et al. 
already took advantage of this to arrive, with very little effort, at the following generalization of the scalar QED
Schwinger formula \eqref{schwingerscal}, which now is restricted to the weak-field limit, but takes an arbitrary number of
photon exchanges between the two nascent particles into account:

\begin{eqnarray}
{\rm Im} {\cal L}^{({\rm all-loop})} (E) \,\,\,\, {\stackrel{\beta\to 0}{\sim}} \,\,\,\,
{\rm Im} {\cal L}^{(1)} (E) 
\,{\rm e}^{\alpha\pi}
\, .
\label{Imall}
\end{eqnarray} 
This result may seem counter-intuitive since it indicates an enhancement while, 
naively, one might think that the attractive force between the two oppositely charged particles
ought to make it more difficult for them to separate out and thereby reduce the pair-creation rate. 
It suggests that the interaction between the particles should not been taken into account before they have turned 
from virtual to real. And indeed, in 1984 Lebedev and Ritus \cite{lebrit} showed that, 
in the tunnelling picture, exactly the same enhancement is obtained by evaluating the Coulomb interaction between
the particles at the critical separation, and interpreting it as an effective lowering of the energy that needs to be drawn from the field in the pair-creation
process. 
With other words, the separating out of the pair along the field lines would have to be seen as
a pure statistical fluctuation unrelated to the equations of motion.

\section{The Vlasov equation}

For purely time-dependent fields, there is the following Vlasov-type equation
describing the time-evolution of the density of created pairs ${\cal N}_{\bf k}$ 
with fixed momentum $\bf k$ \cite{kescm1,kescm2,sbrspt}
\begin{eqnarray}
\frac{d}{dt} (1+ 2 {\cal N}_{\bf k} (t))
&=& \Omega^{(-)} (t) \int_{t_0}^{t} dt' \Bigl[ \Omega^{(-)}(t')
(1+2 {\cal N}_{\bf k} (t') )  \cos( \int_{t'}^{t} dt'' \Omega^{(+)} (t'') ) \Bigr]
\label{vlasov}
\end{eqnarray}
where
\begin{eqnarray}
\Omega^{(\pm)}_k (t) := \frac{\omega_k^2 (t) \pm \omega_k^2 (t_0)}{\omega_k (t_0)}\, ,
\end{eqnarray}
and
\begin{eqnarray}
\omega_k^2 (t) &=& (k_{\parallel} - qA_{\parallel} (t))^2 + {\bf k}_{\perp}^2 + m^2.
\end{eqnarray}
Here $t_0$ is the initial time (usually $-\infty$) and the projections $\parallel,\perp$ refer to the field direction. 

The integro-differential equation \eqref{vlasov} has shown itself to be very amenable to numerical evaluation,
but is usually hopeless for attempts at an exact calculation. An exception is \cite{84,101,147} where an
infinite family of analytic solutions was found related to the well-known solitonic solutions of the Korteweg-de-Vries equation. 
The simplest one has the gauge potential 
\begin{eqnarray}
qA_{\parallel}(t) = \tilde k_{\parallel} - \sqrt{\tilde k_{\parallel}^2+ \frac{2\tilde\omega_0^2}{\cosh^2(\tilde\omega_0 t)}}
\end{eqnarray}
where $\tilde k$ is a fixed reference momentum and $\tilde\omega_0= \sqrt{\tilde{\bf k}^2+m^2}$. Like any purely time-dependent field these ``solitonic'' fields
have non-vanishing total pair creation rates, but there is no pair creation at that particular momentum $\tilde {\bf k}$. 
At intermediate times ${\cal N}_{\tilde{\bf k}} (t)$ is non-zero, but it returns to zero for $t\to\infty$. 
The external field seems to excite the vacuum, but no particles materialize.

\section{How particles are born}

The Dirac-Heisenberg-Wigner formalism, a previously little-used phase-space approach to QED based on a gauge-invariant density operator,
has in the last few years turned out to be capable of providing new insights into the details of the pair-creation process 
\cite{healgi2,hebenstreit,ilmaza,kohlfuerst,dialko}. Here we can only show the following example of a time-like
Sauter field in 1 + 1 dimensions, localized in the space direction:
\bear
\vert e\vert {\bf E}(t,x) = E_0 \, {\rm sech}^2(t/\tau)  {\rm exp}\Bigl(-\frac{x^2}{2\lambda^2}\Bigr) {\bf e}_x
\label{sauter2D}
\ear
where $\tau$, $E_0$ and $\lambda$ are constants. 
A detailed study in momentum space \cite{dialko} shows that there are three time scales $ T_1,T_2,T_3$ involved in the pair-creation process.
The process starts with an early build-up of a narrow peak at $p=0$. Around time $T_1$, a side peak appears. 
Around time $T_2$, the side peak wave packet (now called ``pre-particle'')  starts to follow the classical trajectory.
Around time $T_3$ the central peak has faded away, which concludes the pair-creation process. 
Fig. 6 \footnote{I thank R. Alkofer and C. Kohlf\"urst for providing this figure.} shows snapshots of the particle momentum
density $f_p$, particle space density $f_x$, charge momentum density $q_p$ and charge space density $q_x$ for two different
times. The separating out into two oppositely charged particles is clearly visible.

\vspace{20pt}

\vspace{-10pt}
\begin{figure}[h]
\hspace{-10pt}
{\centering
\includegraphics[scale=.63]{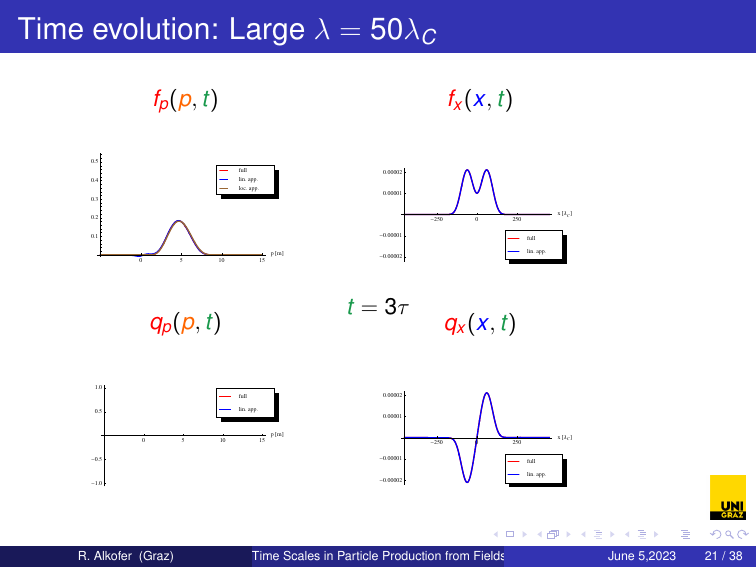}
$\qquad$
\vspace{105pt}
\includegraphics[scale=.63]{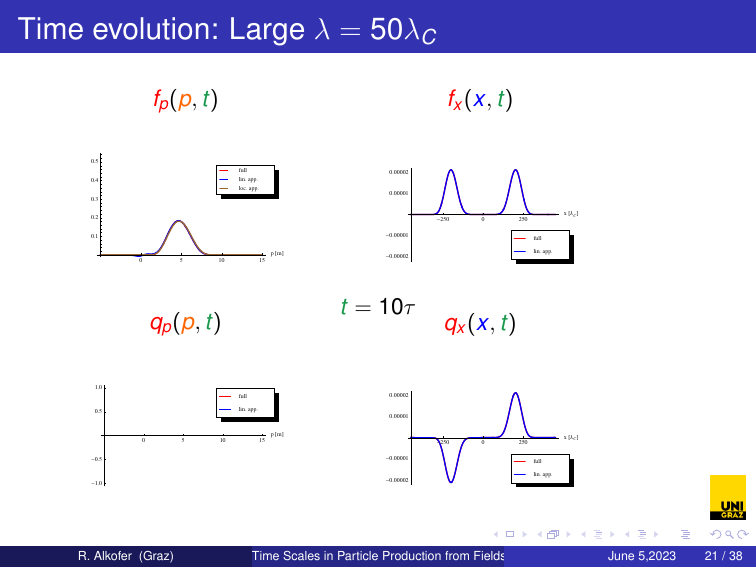}
}
\vspace{-110pt}
\caption{Particle and charge momentum and space densities at two different times for the field \eqref{sauter2D}.}
\end{figure}

\vspace{-20pt}

\section{Summary and outlook}

I have discussed the Sauter-Schwinger pair-creation process with an emphasis on the question of the interaction of the particles in formation
with each other and the external field, an issue that is not easily accessible in the standard formulation of QFT. The recent work of \cite{kohlfuerst,dialko}
on scalar QED demonstrates that the Dirac-Heisenberg-Wigner formalism is capable of providing a detailed space-time picture of this process, 
involving three different time scales and a complex gradual transition from virtual to real particles. 
It would be of great interest to extend these studies to more general theories, in particular to gravity.
Since conservation laws require the interaction between
the nascent particles and the field to be reciprocal, this might also throw new light on the question why straightforward attempts at calculating the contribution of the QFT
vacuum to the cosmological constant so far generally have failed by many orders of magnitude. 

\medskip

\noindent
{\bf Acknowledgements:} I thank Naser Ahmadiniaz and Christian Kohlf\"urst for discussions and correspondence. 

%

\begin{thebibliography}{}
%
%
\bibitem{sauter}
F. Sauter, Z. Phys. {\bf 69} (1931) 742.

\bibitem{schwinger51}
J. Schwinger, 
Phys. Rev. {\bf 82} (1951) 664.

\bibitem{eulhei}
W. Heisenberg and H. Euler,
Z. Phys. {\bf 98} (1936) 714.

\bibitem{LUXE}
H. Abramowicz et al., ``Letter of Intent for the LUXE Experiment'', arXiv:1909.00860 [physics.ins-det].

\bibitem{FACET}
V. Yakimenko et al., ``FACET-II facility for advanced accelerator experimental tests'', 
Phys. Rev. Accelerators and Beams, {\bf 22}, 101301 (2019). 

\bibitem{fikkstt}
A. Fedotov, A. Ilderton, F. Karbstein, B. King, D. Seipt, H. Taya, and G. Torgrimsson,
Phys. Rept. 1010 (2023), arXiv:2203.00019 [hep-ph]. 

\bibitem{rmmhc}
M. Ruf, G.R. Mocken, C. Muller, K.Z. Hatsagortsyan, C.H. Keitel, 
Phys. Rev. Lett. {\bf 102} (2009) 080402, arXiv:0810.4047 [physics.atom-ph].

\bibitem{dugisc}
G.V. Dunne, H. Gies and R. Sch\"utzhold, {\it Phys. Rev.} D {\bf 80} 111301, arXiv:0908.0948 [hep-ph].

\bibitem{keldysh65}
L.~V. Keldysh, {\it Sov. Phys. JETP} {\bf 20} 1307 (1965).

\bibitem{breitz70}
E.~Br\'ezin and C.~Itzykson, Phys.\ Rev.\ D {\bf 2}, 1191 (1970).

\bibitem{narnik70}
N.~B. Narozhny, A.~I. Nikishov, Yad. Fis. {\bf 11} (1970) 1072 [Sov. Journ. Phys. {\bf 11} (1970) 596].

\bibitem{popov72}
V.~S. Popov, Zh. Eksp. Teor. Fiz. {\bf 62} (1972) 1248 [Sov. Phys. JETP {\bf 35} (1972) 569]. 

\bibitem{afalma}
I.K. Affleck, O. Alvarez and N.S. Manton, 
Nucl. Phys. {\bf B 197} (1982) 509-519.



\bibitem{63}
G. V. Dunne and C. Schubert,
Phys. Rev. {\bf D 72} 105004 (2005), arXiv:hep-th/0507174. 

\bibitem{kescm1} Y.~Kluger, J.~M.~Eisenberg, B.~Svetitsky, F.~Cooper and E.~Mottola,
Phys.\ Rev.\ Lett.\ {\bf 67}, 2427 (1991).

\bibitem{kescm2} Y.~Kluger, J.~M.~Eisenberg, B.~Svetitsky, F.~Cooper and E.~Mottola,
Phys.\ Rev.\ D {\bf 45}, 4659 (1992).

\bibitem{sbrspt}
S. Schmidt, D. Blaschke, G. R\"opke, S.~A. Smolyansky,  A.~V. Prozorkevich and V.~D. Toneev,
Int. J. Mod. Phys. E {\bf 7} (1998) 709, arXiv: hep-ph/9809227.

\bibitem{healgi}
F.~Hebenstreit, R.~Alkofer and H.~Gies, 
Phys.\ Rev.\ D {\bf 78}, 061701 (2008), arXiv:0807.2785.

\bibitem{wigner32}
E. Wigner, Phys. Rev. {\bf 40} 749 (1932). 

\bibitem{vagyel}
D. Vasak, M. Gyulassy and H.-T. Elze,
Ann. Phys. {\bf 173} 462 (1987). 


\bibitem{bigora}
I. Bialynicki-Birula, P. Góritzki and J. Rafelski,
Phys. Rev. D {\bf 44} 1825 (1991).

\bibitem{ochhei}
S. Ochs and U. Heinz,
Ann. Phys. {\bf 266}, 351 (1998). 


\bibitem{healgi2}
F.~Hebenstreit, R.~Alkofer and H.~Gies, 
Phys. Rev. D {\bf 82}, 105026 (2010),  arXiv:1007.1099 [hep-ph].

\bibitem{hebenstreit}
F. Hebenstreit, ``Schwinger effect in inhomogeneous electric fields'', PhD thesis, Graz University, 2011, arXiv:1106.5965 [hep-ph]. 

\bibitem{ilmaza}
F. Hebenstreit, A.Ilderton, M. Marklund and J. Zamanian,  Phys. Rev. D {\bf 83} (2011) 065007, arXiv:1011.1923 [hep-ph].

\bibitem{kohlfuerst}
C. Kohlf\"urst,
``Electron-positron pair production in inhomogeneous electromagnetic fields'',
PhD thesis, Graz University, 2015, arXiv:1512.06082 [hep-ph]. 

\bibitem{dialko}
M.~Diez, R.~Alkofer and C.~Kohlfürst,
Phys. Lett. B {\bf 844} (2023) 138063, arXiv: 2211.07510 [hep-ph]. 

\bibitem{feynman1950}
R. P. Feynman, Phys. Rev. {\bf 80} (1950) 440.

\bibitem{kimpage1} 
S.~P.~Kim and D.~N.~Page, Phys.\ Rev.\ D {\bf 65}, 105002 (2002), hep-th/0005078.

\bibitem{kimpage2} 
S.~P.~Kim and D.~N.~Page, Phys.\ Rev.\ D {\bf 73}, 065020 (2006), hep-th/0301132.

\bibitem{kimpage3} 
S.~P.~Kim and D.~N.~Page, Phys.\ Rev.\ D {\bf 75}, 045013 (2007), hep-th/0701047. 

\bibitem{nikishov}
A.~I. Nikishov, Nucl. Phys. B {\bf 21} (1970) 346.

\bibitem{giekli}
H.~Gies and K.~Klingm\"uller, Phys.\ Rev.\ D {\bf 72}, 065001 (2005), hep-ph/0505099.

\bibitem{64}
G. V. Dunne, Q.-h. Wang, H. Gies and C. Schubert,  
{\it Phys. Rev.} D {\bf  73} 065028 (2006),  hep-th/0602176.

\bibitem{schsch-schwinger}
C. Schneider and R. Sch\"utzhold, 
JHEP (2016) 164,  arXiv:1407.3584 [hep-th]. 

\bibitem{lebrit} 
S.L. Lebedev and V.I. Ritus, 
Zh. Eksp. Teor. Fiz. {\bf 86} (1984) 408 [JETP {\bf 59} (1984) 237].

\bibitem{84}
S. P. Kim and C. Schubert, 
Phys. Rev. D {\bf 84} 125028 (2011), arXiv:1110.0900 [hep-th].


\bibitem{101}
 A. Huet, S. P. Kim and C. Schubert, 
 Phys. Rev. D {\bf 90} (2014) 125033, arXiv:1411.3074 [hep-th].


\bibitem{147}
N. Ahmadiniaz, A.M. Fedotov, E.G. Gelfer, S. P. Kim and C. Schubert,
Phys. Rev. D {\bf 108}, 036019 (2023); arXiv:2205.15946 [hep-th]. 


%
%
\end{thebibliography}
%
%

\end{document}